\documentclass{article}
\usepackage{dcase2025_techrep,amsmath,url,times,booktabs, tabularx}
\usepackage{graphicx}
\usepackage{soul,color}
\usepackage{caption}
\usepackage{subcaption}
\usepackage{amssymb}
\usepackage{hyperref}
\usepackage{parskip}
\usepackage{setspace}

\def\etal{\emph{et al}.\ }

\title{Spatial and Semantic Embedding Integration for Stereo Sound Event Localization and Detection in Regular Videos}

\twoauthors
  {Davide Berghi\sthanks{This research was funded by EPSRC-BBC Prosperity Partnership `{AI4ME}: Future personalised object-based media experiences delivered at scale anywhere' (EP/V038087/1). 
For the purpose of open access, the authors have applied a Creative Commons Attribution (CC BY) license to any Author Accepted Manuscript version arising. 
Data supporting this study is available from \url{https://zenodo.org/records/15559774}}}
    {CVSSP, University of Surrey, U.K.\\
     davide.berghi@surrey.ac.uk}
  {Philip J.\,B.\ Jackson}
    {CVSSP, University of Surrey, U.K.\\
     p.jackson@surrey.ac.uk}

\begin{document}

\ninept
\maketitle

\begin{sloppy}

\begin{abstract}

This report presents our systems submitted to the audio-only and audio-visual tracks of the DCASE2025 Task 3 Challenge: Stereo Sound Event Localization and Detection (SELD) in Regular Video Content.
SELD is a complex task that combines temporal event classification with spatial localization, requiring reasoning across spatial, temporal, and semantic dimensions. The last is arguably the most challenging to model. Traditional SELD architectures rely on multichannel input, which limits their ability to leverage large-scale pre-training due to data constraints.
To address this, we enhance standard SELD architectures with semantic information by integrating pre-trained, contrastive language-aligned models: CLAP for audio and OWL-ViT for visual inputs. These embeddings are incorporated into a modified Conformer module tailored for multimodal fusion, which we refer to as the Cross-Modal Conformer.
Additionally, we incorporate autocorrelation-based acoustic features to improve distance estimation. We pre-train our models on curated synthetic audio and audio-visual datasets and apply a left-right channel swapping augmentation to further increase the training data.
Both our audio-only and audio-visual systems substantially outperform the challenge baselines on the development set, demonstrating the effectiveness of our strategy. Performance is further improved through model ensembling and a visual post-processing step based on human keypoints. Future work will investigate the contribution of each modality and explore architectural variants to further enhance results.
 
\end{abstract}


\section{Introduction}
\label{sec:intro}

Sound Event Localization and Detection \cite{Adavanne:2019:SELDnet} is a combined task that integrates sound event detection (SED) \cite{Adavanne:2017:sed} and sound source localization (SSL) \cite{Adavanne:2018:DOA}. The goal is to identify active sound events from predefined target classes, track their temporal activity, and estimate their spatial positions.
SELD systems are crucial for a wide range of real-world applications, including human-robot interaction \cite{Xinyuan:2023:AVcrossAtt}, security monitoring, and immersive media production \cite{Berghi:2024:forecasterFlexOBM}. 

Since 2019, SELD has been a dedicated task in the Detection and Classification of Acoustic Scenes and Events (DCASE) Challenge.
Over successive editions, the task has evolved to include increasingly complex scenarios, such as detecting moving sound sources \cite{politis:2020:DCASE}, ignoring external interfering sounds \cite{politis:2021:DCASE}, incorporating visual input to enable multimodal SELD in 360$^{\circ}$ videos \cite{Shimada2023STARSS23AA}, and estimating source distance \cite{Diaz-Guerra:2024:seldBaseline24}.
In this year's edition, the challenge has shifted from the traditional 4-channel first-order ambisonics (FOA) and microphone array (MIC) formats to stereo SELD using conventional frontal video content, i.e., leveraging only left and right audio channels and perspective video. This format is better aligned with the requirements of conventional media content.
In stereo SELD the DOA prediction is limited to the azimuth angles in the range [-90$^{\circ}$, 90$^{\circ}$], as elevation 
is ill-defined when using
just two horizontally arranged channels, and distinguishing between front and back sources is inherently ambiguous. Source distance estimation, introduced in the previous edition, remains part of the task. Additionally, in the audio-visual track, a new subtask has been introduced: predicting whether sound sources are onscreen or offscreen.

SELD is a non-trivial task, as it requires reasoning across spatial, temporal, and semantic dimensions: spatial information gives cues for direction and distance estimation;
temporal information marks movements and onsets/offsets of sound source activity; semantic information identifies objects, their relations and likely behaviors.
Recent advances in large language models (LLMs) \cite{Brown:2020:GPT}, vision-language models (VLMs) \cite{Radford:2021:clip}, and audio-language models (ALMs) \cite{Huang:2024:AudioGPT} demonstrate that language is a powerful lens for enabling semantic understanding and complex reasoning in relation to media content. 
Building on this, we posit that leveraging language-aligned models can enhance a SELD model’s semantic reasoning and hence indirectly benefit spatial and temporal reasoning. 
Traditionally, spatial localization relies on multichannel audio, allowing models to infer source positions through inter-channel time and level differences. However, this dependence limits the use of large-scale pre-training datasets, as language alignment typically requires extensive pre-training on large and diverse datasets.
Therefore, we extend our SELD architecture by integrating two existing language-aligned models: CLAP \cite{wu:2023:clap_laion} for the audio modality and OWL-ViT \cite{Minderer:2022:owl_vit} for the visual modality. Specifically, we extract audio and visual embeddings using their respective pretrained encoders and combine them with SELD embeddings obtained from a CNN-Conformer backbone, a widely adopted architecture in SELD research \cite{Wang:2023:ACS,Berghi:2024:ICASSP24,Xue:2023:resnetConf}. We argue that the embeddings from CLAP and OWL-ViT are semantically rich and provide complementary information to the task.
To fuse these embeddings, we employ an adapted version of the Conformer architecture \cite{Gulati2020ConformerCT}, which we refer to as the Cross-Modal Conformer (CMC)\@. This module is used to integrate intra-modal embeddings from different sources (e.g., the SELD audio encoder and CLAP) and inter-modal embeddings (i.e., the combined audio representation and the visual embedding from OWL-ViT).
This architecture is outlined in Sec.~\ref{sec:model}.


\begin{figure}[bt]
\centering
  \includegraphics[width=\columnwidth]{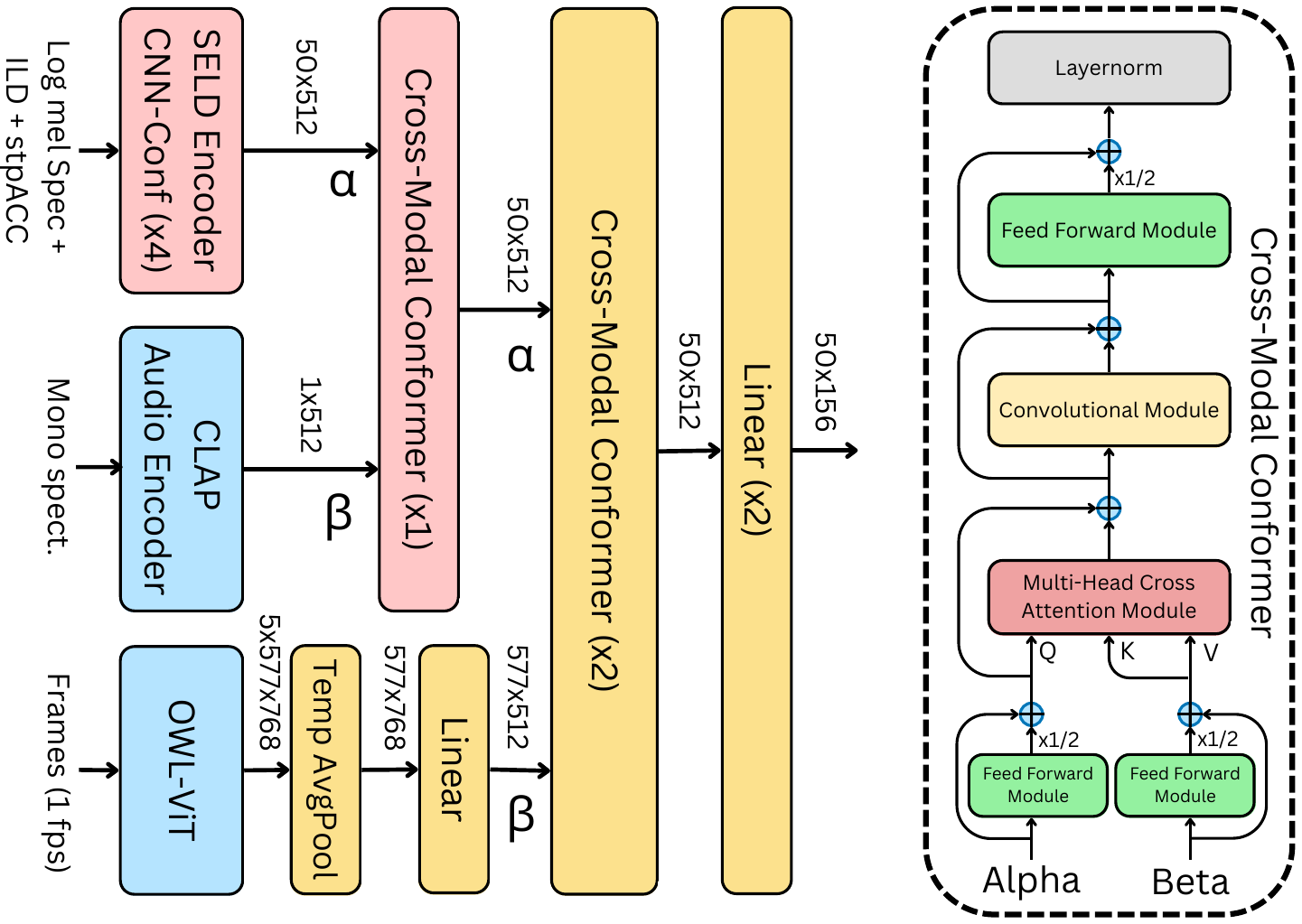} \\ [1ex]
\caption{Proposed audio-visual model (left). Blue blocks are frozen, red ones are pre-trained on the audio 5k and audio-visual 2k datasets, yellow on audio-visual 2k dataset only. 
Cross-Modal Conformer (CMC) architecture adapting the original Conformer \cite{Gulati2020ConformerCT} to process two generic modalities ``Alpha'' $\alpha$ and ``Beta'' $\beta$ (right).}
\label{fig:model} \vspace{-4mm}
\end{figure}

This report describes the two audio systems we submitted to track A, and the four audio-visual systems we submitted to track B of the challenge. To tackle the distance estimation subtask, we partnered common SELD input features with short-term power of the autocorrelation stpACC features \cite{Berghi:2025:distanceFeat} in Sec.~\ref{sec:features}. 
We synthesized and carefully curated large audio and audio-visual datasets used to pre-train our models, described in Sec.~\ref{sec:preproc}.
Our audio-visual work includes a visual post-processing step based on human keypoint detection and model ensembling, which is described in Sec.~\ref{sec:exp} on experiments.

\section{Proposed Architecture}
\label{sec:model}

The proposed model incorporates a primary SELD encoder that extracts SELD embeddings from multichannel input features. These embeddings are then integrated with the CLAP audio embedding through a cross-modal cross-attention mechanism. We adapted the Conformer architecture to accommodate inputs from different modalities.
The resulting audio representation is subsequently fused with a visual embedding extracted using OWL-ViT leveraging a second cross-modal Conformer. Finally, a feed-forward module composed of two linear layers predicts multi-ACCDDOA vectors modeling up to $N{=}3$ tracks \cite{Krause:2024:seldDistance}, including on- and off-screen activity predictions as in the challenge baseline.
The model is trained using class-wise Auxiliary Duplicating Permutation Invariant Training (ADPIT) loss \cite{Shimada:2022:multiACCDOA,cao:2020:EIN,cao:2021:EINv2}.

\subsection{SELD Encoder}
\label{subsec:seld_enc}

For the SELD encoder, we adopted a CNN-Conformer architecture, as it is widely adopted in SELD research \cite{Wang:2023:ACS,Berghi:2024:ICASSP24,Xue:2023:resnetConf}. It consists of a CNN encoder followed by a Conformer module \cite{Gulati2020ConformerCT}.
The CNN encoder processes stereo input features, with an input of shape $C_\mathrm{in}{\times} T_\mathrm{in}{\times} F_\mathrm{in}$. 
Where, $C_\mathrm{in}$ represents the number of feature channels, while $T_\mathrm{in}$ and $F_\mathrm{in}$ the number of temporal and frequency bins, respectively.
The CNN encoder comprises four convolutional blocks with residual connections, each containing two 3$\times$3 convolutional layers, BN \cite{ioffe:2015:BN}, ReLU activation, and Avg pooling with a stride of 2, halving the temporal and frequency dimension at each block. The resulting tensor of shape $512\times T_\mathrm{in}/16\times F_\mathrm{in}/16$ is reshaped and frequency Avg pooling is applied to achieve a $T_\mathrm{in}/16\times 512$ embedding. 
$T_\mathrm{in}$ is chosen so that $T_\mathrm{in}/16$ matches the label frame rate (10 labels\,/\,sec).
A Conformer module with four layers and eight attention heads processes this embedding, using depthwise convolutions with kernel size of 51.

\subsection{Cross-Modal Conformer}
\label{subsec:cm_conf}

The CMC is an adapted version of the Conformer architecture proposed by Gulati \etal \cite{Gulati2020ConformerCT}. A schematic representation is shown on the left side of Figure\,\ref{fig:model}, where a generic modality ``Alpha'' ${\in}\,\mathbb{R}^{T_\alpha \times d_k}$ is combined with another modality ``Beta'' ${\in}\,\mathbb{R}^{T_\beta \times d_k}$ \cite{Xinyuan:2023:AVcrossAtt}. Each sub-module within the architecture retains the structure of the original Conformer modules, with the key modifications being two initial feed-forward layers that independently process the two modalities in parallel. Additionally, the standard multi-head self-attention mechanism is replaced by a multi-head cross-attention module, where queries are derived from modality Alpha and keys/values from modality Beta.
We utilize two CMC blocks in our model: the first fuses SELD embeddings with semantically rich CLAP embeddings; the second combines the resulting audio representation with semantically and spatially rich visual embeddings extracted by OWL-ViT. We found that a single layer sufficed for audio fusion, while two layers yielded the best performance for audio-visual fusion.
Consistent with the Conformer module of the SELD encoder, each CMC employs eight attention heads and depthwise convolutions with a kernel size of 51.

\subsection{Video Embedding Extraction}

Since SELD involves spatial, semantic, and temporal reasoning, our previous works \cite{Berghi:2024:ICASSP24,Berghi:2024:DCASE24techRep} employed ResNet50 \cite{He:2016:resnet} to extract visual features from each individual video frames, capturing the temporal dimension across time. To obtain per-frame visual embeddings, a $7{\times}7$ average pooling operation was applied, resulting in a single feature vector for each frame.
However, we now argue that this approach degrades spatial resolution, limiting the model’s ability to utilize fine-grained spatial cues. In contrast, other works, such as \cite{hong:2024:MVAnet} or this year’s baseline system, apply average pooling across the channel dimension of the ResNet50 output, preserving the $7{\times}7$ spatial layout. Yet, we believe that this alternative sacrifices semantic richness, as pooling across channels degrades the learned feature representations.

Effectively managing time, space, and channel information in a unified framework is non-trivial. Nevertheless, we argue that temporal dynamics can be effectively captured by the SELD encoder itself. As such, the visual processing branch should be optimized to better leverage semantic and spatial information.
To this end, we sacrifice temporal granularity in the visual stream. We replaced ResNet50 with OWL-ViT \cite{Minderer:2022:owl_vit}, a contrastive, language-aligned model, like CLAP, but specifically trained for visual grounding tasks like object detection. As a result, OWL-ViT produces visual embeddings that are both semantically and spatially rich.
To retain some temporal context without incurring high computational costs, we sample video frames at 1 fps. The resulting embeddings are then aggregated via temporal average pooling. All individual ViT token embeddings are maintained to preserve spatial and semantic information.

OWL-ViT requires square input frames of size $768{\times}768$. To avoid distorting the original rectangular frames in the dataset, we explored two pre-processing strategies during preliminary experiments: letterboxing the frames with black bars to preserve the aspect ratio, and applying a non-linear spatial transformation that preserves object proportions in the central region while progressively stretching the top and bottom areas. This second approach is based on the assumption that most sound events occur near the center of the frame.
The non-linear re-framing method yielded slightly better performance, and we therefore integrated it into our pre-processing pipeline.
Visual tokens are extracted from $32{\times}32$ pixel patches, resulting in 576 tokens per frame, plus an additional classification token. Each token embedding has a dimensionality of 768. To align with the model's architecture, we apply a linear projection to reduce this dimensionality to $d_k{=}512$. These OWL-ViT embeddings serve as key/value pairs (i.e., modality ``Beta'') in the second CMC.

\section{Acoustic Input Features}
\label{sec:features}

Since the left and right audio channels in the dataset are arithmetically derived from FOA signals, rather than captured by two physically separated microphones, they should not present inter-channel time or phase differences. 
So, we adopted the inter-channel level difference (ILD) as the primary spatial feature for the SELD encoder, alongside log mel spectrograms computed independently from each channel. ILD features are calculated as the ratio of the squared magnitudes of the short-time Fourier transforms (STFTs) of the two channels, and subsequently mapped into the log mel domain: 
\begin{equation}
\mathbf{ILD}(m,t) = \log\left( \mathbf{H}_{\mathrm{mel}}  \frac{|\mathbf{L}(f,t)|^2+ \epsilon}{|\mathbf{R}(f,t)|^2 + \epsilon} \right),
\end{equation}
where $\mathbf{L}(f,t)$ and $\mathbf{R}(f,t)$ are the STFTs of the left and right channels, respectively, $\mathbf{H}_{\mathrm{mel}}$ is the mel filter bank, $\epsilon$ is a small constant to avoid division by zero and instability, and $m$ in the mel frequency index.
To further support the model in the distance estimation subtask, we include short-term power of the autocorrelation (stpACC) features, as proposed in \cite{Berghi:2025:distanceFeat}. 
The final concatenation of acoustic input features has a channel dimension of $C_\mathrm{in}{=}4$.

\section{Pre-processing and Data Augmentation}
\label{sec:preproc}

We pre-trained our model on synthetic data while keeping the CLAP audio encoder and OWL-ViT weights frozen. Synthetic FOA audio was generated using SpatialScaper \cite{Roman:2024:spatialScaper}, which convolves FSD50K sounds \cite{fonseca2022FSD50K} with RIRs from various datasets \cite{politis:2020:DCASE,orhun_olgun_2019_2635758,mckenzie2021dataset,defferrard2016fma,gotz2021dataset,chesworth2024room,schneiderwind2019data}. We created 5,000 one-minute FOA clips averaging 18 events per clip (std=6), and 500 additional clips for validation. Background noise levels were sampled from a normal distribution (mean=\,–65\,dB, std=15) to encourage robustness. We overrepresented ``Knock'' and ``Bell'' sounds due to their detection difficulty.
Using the provided stereo SELD generator, we split each FOA clip into twelve 5-second segments (i.e., with hop size=5s), filtered out silent ones, and applied four random FOA rotations to each segment before extracting the stereo sounds, resulting in $\sim$150k training clips. We'll refer to this synthetic dataset as ``audio 5k'', as it is derived from 5,000 FOA files. Models were then fine-tuned on the real content of the stereo STARSS dataset.

For the audio-visual model, we generated an additional 2,000 FOA clips with SpatialScaper and synthesized corresponding videos using SELDVisualSynth \cite{roman2025generating}. To enrich the quality of the visual output, we manually curated and extracted class-relevant images from Flickr30k \cite{Plummer:2015:flickr}. We also selected images of doors from DoorDetect \cite{Arduengo_2021}, which have been employed for ``door open/close'' and ``knock'' sounds. 
Furthermore, we created additional synthetic foreground images for each sound class using NitroFusion \cite{chen:2024:nitrofusion}.  
Instead of employing the background canvases provided with SELDVisualSynth which consists primarily of outdoor environments, we adopted the +2,000 environments of the 360-Indoor dataset \cite{Chou:2020:360indoor}.
We also implemented a soft cross-fade between background canvases and foreground object tiles to remove strong artificial edges from the pre-training dataset.
Figure\,\ref{fig:synthFrames} shows example frames from the dataset.
We initialized the SELD encoder and audio CMC with weights from the audio-only pre-training, and continued training on the +60,000 audio-visual clips generated following this approach. We refer to this dataset here as ``audio-visual 2k''.

To further increase the size of the training dataset and model robustness, we adapted augmentations to the stereo scenario, the audio-channel swap (ACS) \cite{Wang:2023:ACS} and video pixel swap (VPS) \cite{Berghi:2024:ICASSP24,Jiang:2024:AVseld,Roman:2024:ehnancedAVseld}. Thus, we swapped the left and right audio channels, and flipped the corresponding video frames to double the size of the training data.
The synthetic data augmented with such methods gave over 410h of stereo audio data and an additional 168h of audio-visual data.

\begin{figure}[tb]
\begin{minipage}{0.49\columnwidth}
\includegraphics[width=\columnwidth]{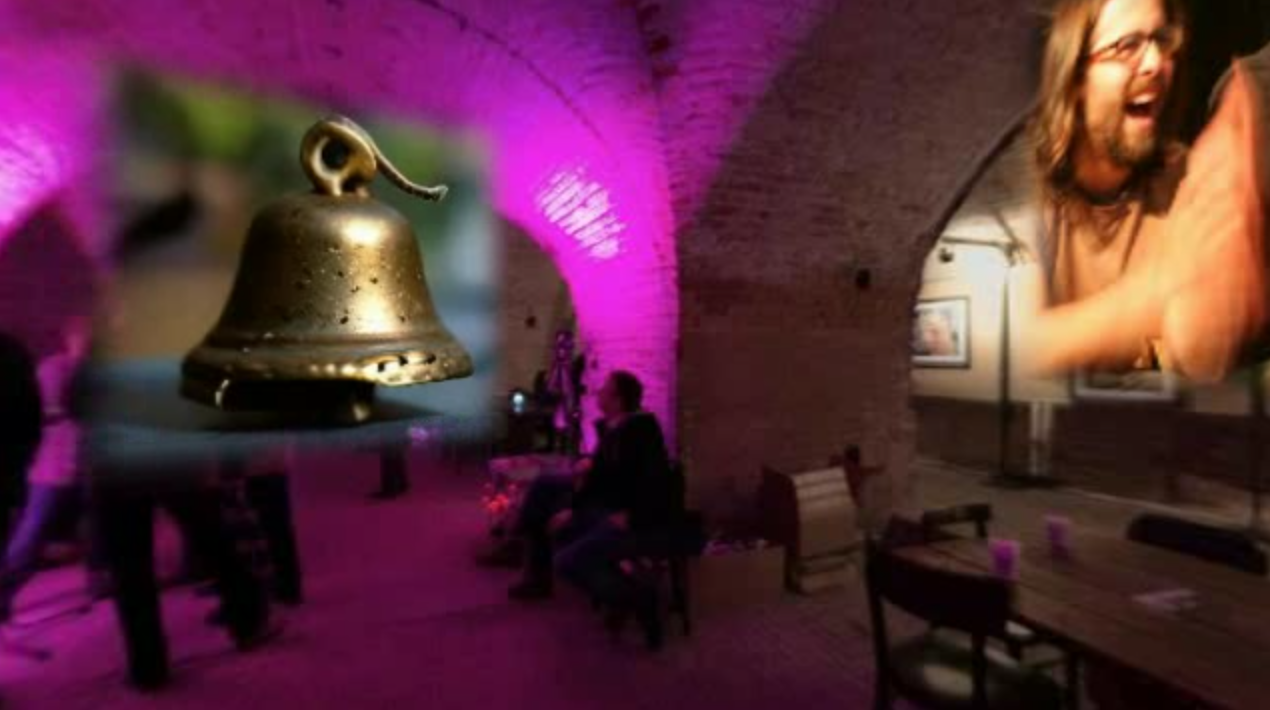}
\vspace{-3mm}
\end{minipage}\hfill
\begin{minipage}{0.49\columnwidth}
\includegraphics[width=\columnwidth]{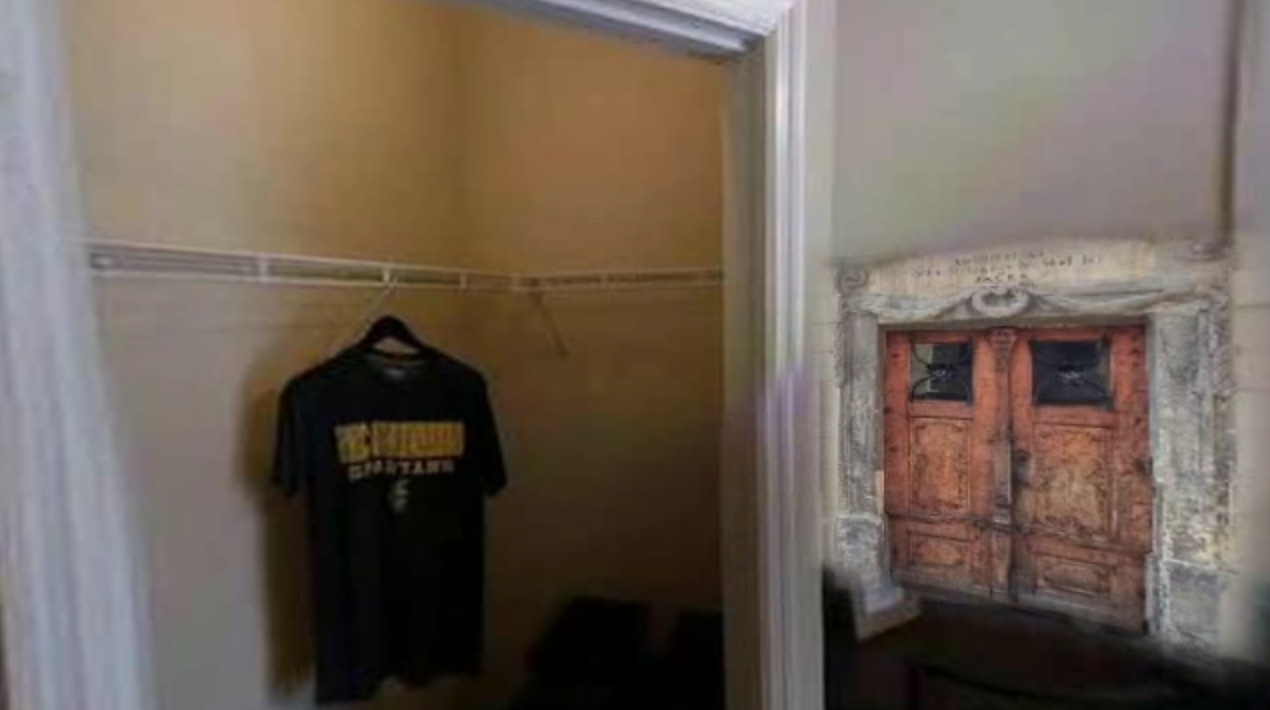} 
\vspace{-3mm}
\end{minipage}

\begin{minipage}{0.49\columnwidth}
\includegraphics[width=\columnwidth]{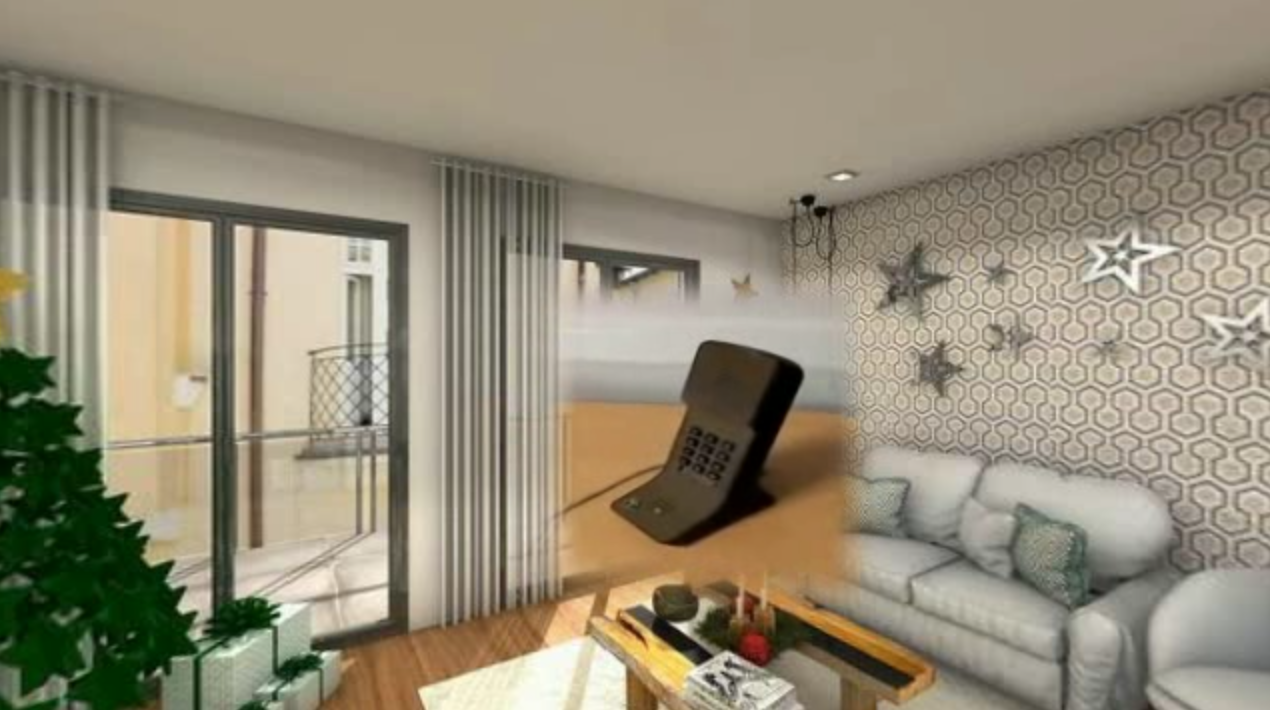}
\end{minipage}\hfill
\begin{minipage}{0.49\columnwidth}
\includegraphics[width=\columnwidth]{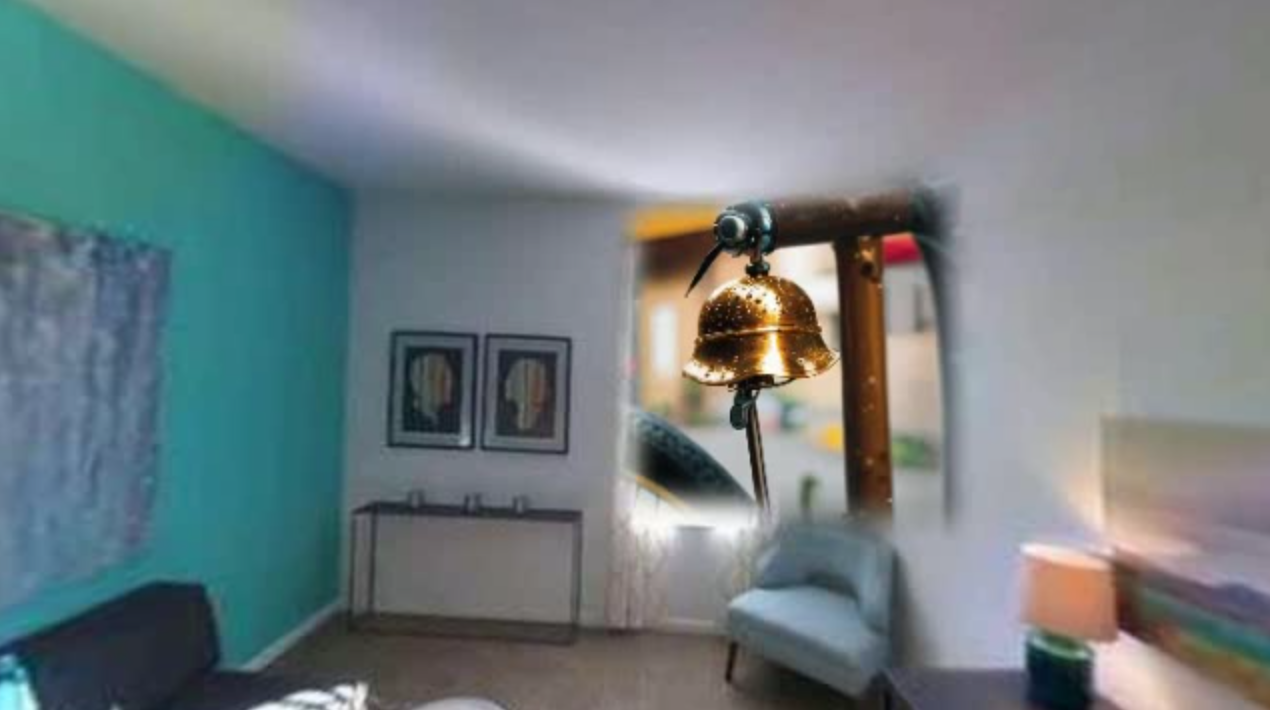} 
\end{minipage}

\captionof{figure}{Examples of synthetic scenes (clockwise from top left) in 
a restaurant, a walk-in wardrobe, a hotel room, a first-floor flat.
The foreground tiles (bell*, laughter, door*, bell*, telephone) are applied with a soft cross-fade to avoid strong edges. *~Generated with NitroFusion \cite{chen:2024:nitrofusion}.}
\label{fig:synthFrames}
\end{figure}

\section{Experiments}
\label{sec:exp}

\subsection{Implementation Details}

Audio spectrograms were generated via STFT with a 512-point Hann window and 150-sample hop size. With a sampling rate of 24\,kHz, this produces 800 temporal bins for the 5-second input clips. We used 64 mel bins for our spectrograms.
For stpACC features, we applied an STFT with a 1014-point Hann window.
This ensures that the autocorrelation covers delays up to approximately 20\,ms after the direct sound. We then downsample the time-lag dimension by a factor of 8 to achieve 64 bins and allow concatenation with the other features. More details about stpACC can be found in \cite{Berghi:2025:distanceFeat}.
The SELD encoder and the first CMC are pre-trained on the synthetic audio 5k dataset for 100 epochs. Subsequently, the full audio-visual model is further trained on the synthetic audio-visual 2k dataset for other 80 epochs. Finally, the model is fine-tuned on the training partition of the real development set for 80 epochs, selecting the best based on the highest F1 score on the test partition.
Models are trained with batch size 32 using Adam optimizer. After the first 30 epochs, the learning rate is reduced by 5\% per epoch.

Our audio-only model is similar to the audio-visual one, but without the OWL-ViT branch. Additionally, the second CMC is replaced by a classic Conformer unit to match the depth of the audio-visual model. 
Since the stereo viewing angle in the dataset is randomly sampled from the full 360$^{\circ}$ frames of the STARSS23 dataset \cite{Shimada2023STARSS23AA}, the number of off-screen sound events is significantly higher than that of on-screen events. As a result, we observed that the model tends to consistently predict ``off-screen'', effectively suppressing on-screen predictions. This behavior allows the model to reach a stable $\sim$80\% accuracy, reflecting a bias toward the dominant class.
To address this imbalance, we train a second model using a modified loss function, in which the binary cross-entropy (BCE) loss for on/off-screen predictions is multiplied by a factor of 4.0 whenever the ground truth label is on-screen. 

\subsection{Visual Post-Processing}

Inspired by the work of Jiang \etal \cite{Jiang:2024:AVseld}, we implemented a visual post-processing step based on human keypoint detection. Specifically, we process the video frames using YOLOv11-Pose \cite{Jocher:2023:yolo11pose} to extract keypoints. We then map specific sound event classes to corresponding human keypoints: female speech, male speech, and laughter are associated with the nose keypoint; clapping with the average position of the left and right wrists; and footsteps with the average position of the left and right ankles.
If the predicted direction of arrival (DOA) for any of these classes falls within 20$^{\circ}$ of the corresponding keypoint’s direction, we replace the predicted position with that of the keypoint.
We found that this method resulted in a slight decrease in spatial accuracy. This drop may be attributed to events being mistakenly linked to nearby individuals who are actually located in the opposite hemisphere but with similar lateral position, or distortions introduced by the perspective projection, which causes a non-linear mapping between pixel coordinates and DOA values.
For this reason, we ultimately chose not to apply this spatial refinement technique to the DOA predictions. However, we do use it to set the on/off-screen label to ``on-screen'' for all sound event predictions associated with a visible human. As a result, this post-processing step only affects the on/off-screen accuracy and the $\mathrm{F_{\leq 20^{\circ}/1/on}}$ metrics.

\subsection{Ensemble Strategy}

To further improve performance, we employ an ensemble strategy that combines predictions from multiple systems.
At each time frame, a sound event is considered active only if it is detected by at least two systems, and their predicted DOAs are within 20$^{\circ}$ of each other. For such events, the ensemble DOA is computed as the average of the DOAs from all systems that detected the event.
There are two exceptions to this rule: for the ``Bell'' and ``Knock'' classes, a detection from even a single system is sufficient to consider the event active, as in
\cite{Berghi:2024:DCASE24techRep}.
Similarly, for the on/off-screen classification, if at least one system predicts the event as on-screen, the ensemble output is also considered on-screen.

\begin{table}[bt]
\caption{Results on the development set of stereo STARSS23 \cite{Shimada2023STARSS23AA}: $\mathrm{F_{\leq 20^{\circ}/1}}$ (F1) [\%], $\mathrm{F_{\leq 20^{\circ}/1/on}}$ (F1o) [\%], Direction Of Arrival Error (DOAE) [°], On/Off Accuracy (Acc) [\%]. 
The models audio-only (AO) or audio-visual (AV); variants with weighted loss (+W), visual post-processing (+P); Ensemble (Ens.)\ combines (1),(2),(3) and (4). Submitted challenge models are denoted as $^{\mbox{\scriptsize a1,a2,b1,b2,b3,b4}}$. 
}
\vspace{-2mm}
\centerline{\begin{tabular}{l|c|c|c|c|c} \hline
\textbf{Model} & F1\,$\uparrow$ & F1o\,$\uparrow$ & DOAE\,$\downarrow$ & RDE\,$\downarrow$ & Acc\,$\uparrow$ \\ \hline
Baseline AO 
& 22.8 & - & 24.5 & 41.0 & -  \\
Baseline AV 
& 26.8 & 20.0 & 23.8 & 40.0 & 80.0  \\ \hline
(1) AO$^{\mbox{\scriptsize a1}}$ 
& 45.7 & - & 15.0 & 31.0 & - \\
(2) AO$^{\mbox{\scriptsize a2}}$ 
& 46.0 & - & 15.2 & 30.8 & - \\ \hline
(3) AV 
& 44.4 & 34.0 & 15.6 & 30.4 & 80.5  \\
(3.1) AV+P$^{\mbox{\scriptsize b1}}$ 
& 44.4 & 34.4 & 15.6 & 30.4 & 80.5  \\
(4) AV+W 
& 45.5& 35.4 & 15.2 & 32.2 & 80.8  \\
(4.1) AV+W+P$^{\mbox{\scriptsize b2}}$ 
& 45.5 & 35.7 & 15.2 & 32.2 & \textbf{81.0}  \\
(5) Ens.$^{\mbox{\scriptsize b3}}$ 
& \textbf{48.0} & 37.3 & \textbf{14.0} & \textbf{29.3} & 80.8  \\
(5.1) Ens.+P$^{\mbox{\scriptsize b4}}$ 
& \textbf{48.0} & \textbf{37.5} & \textbf{14.0} & \textbf{29.3} & 80.8  \\ \hline
\end{tabular}
}
\label{tab:results}
\vspace{-2mm}
\end{table}

\subsection{Results}

The results for our audio-only (AO) and audio-visual (AV) systems are presented in Table\,\ref{tab:results}.
The two AO systems are identical but were trained with slightly different initial learning rates. We chose to report and submit both, as they both contribute to improving the ensemble performance.
System (4) uses the same AV model as system (3), but with the loss function that weights the on/off-screen predictions more heavily when the event is on-screen.
The ensemble model (5) combines predictions from systems (1), (2), (3) and (4).
Finally, applying visual post-processing to systems (3), (4) and (5) results in the enhanced versions: (3.1), (4.1) and (5.1), respectively.

Both AO and AV systems substantially outperform the challenge baselines.
The AO systems achieved slightly better $\mathrm{F_{\leq 20^{\circ}/1}}$ and DOAE scores compared to the base AV systems (3) and (4), which also need to predict on/off-screen labels, adding an extra layer of complexity.
Using a weighted loss function for on/off-screen classification gave only a minor gain in overall on/off accuracy. 
However, it led to a 1.4 percentage point gain in $\mathrm{F_{\leq 20^{\circ}/1/on}}$\@.
The visual post-processing step, which only modifies on-screen predictions, brought small gains in $\mathrm{F_{\leq 20^{\circ}/1/on}}$\@. 
$\mathrm{F_{\leq 20^{\circ}/1}}$, DOAE and RDE remained unchanged as post-processing has no effect on localization and class predictions.
Finally, we observed RDE improvements of approximately 10 percentage points compared to the baselines. 
These gains may be attributed in part to the inclusion of stpACC features, which enhance the model’s ability to estimate distance.

\section{Conclusion}
\label{sec:concl}

This technical report describes the systems submitted to Task 3 of the DCASE 2025 Challenge.
Our approach leverages a model that integrates semantically rich feature embeddings from CLAP and OWL-ViT, fused through an adapted Conformer architecture. The model is pre-trained on large, curated synthetic audio and audio-visual datasets.
All submitted systems outperform both the audio-only and audio-visual baselines on the test partition of the development set. Additional performance gains are achieved through model ensembling and visual post-processing.


\bibliographystyle{IEEEtran}
\setlength{\parskip}{0pt plus0pt minus0pt} 
\setlength{\parindent}{0pt}
\setlength{\itemsep}{0pt}
{ \setstretch{0.975}
\bibliography{refs}
}
%
%
%
%
%
%
%
%
%

\end{sloppy}
\end{document}